%% file: main.tex
\documentclass{article}
\usepackage[a4paper, margin=0.72in]{geometry}
\usepackage{multicol}
\usepackage{amsmath}
\usepackage{graphicx}
\usepackage{caption}
\usepackage{subcaption}
\usepackage{float}
\usepackage{enumitem}
\usepackage[font=small,labelfont=bf]{caption}
\title{\textbf{Relativistic Tight-Binding Model for Hexagonal Lattice: Application to Graphene.}}
\author{Rohin Sharma\textsuperscript{1,*},  Amit Shrestha\textsuperscript{2}, Katsuhiko Higuchi\textsuperscript{3}, Masahiko Higuchi\textsuperscript{4}, Dipendra B. Hamal\textsuperscript{5}}
\date{}
\begin{document}
\maketitle
\begin{center}
\emph{\textsuperscript{1} Department of Physics, University of South Florida, Tampa, FL 33620, USA.}\\
\emph{\textsuperscript{2} Institute of Materials Chemistry and Engineering, Kyushu University, Japan.}
\\
\emph{\textsuperscript{3} Graduate School of Advanced Science of Matter, Hiroshima University, Higashi-Hiroshima 739-8527, Japan.}
\\
\emph{\textsuperscript{4} Department of Physics, Faculty of Science, Shinshu University, Matsumoto, Nagano 390-8621, Japan.}
\\
\emph{\textsuperscript{5} Department of Physics, Kathmandu University, Nepal.}
\\
\textsuperscript{*}Corresponding email: rohin@usf.edu
\end{center}
\noindent
\textbf{ABSTRACT:} A non-perturbative relativistic Tight-Binding (TB) approximation method applicable to crystalline material immersed in a magnetic field was developed and tested for crystalline silicone in 2015. To apply this method to any material in a magnetic field, the relativistic hopping integrals in the absence of magnetic field must be calculated. In this study, we formulate the relativistic tight-binding model for hexagonal lattice structure and carry out the electronic structure calculations for graphene in absence of magnetic field. The Hamiltonian and overlap matrix is constructed considering the nearest neighbouring atomic interactions between the $s$ and $p$ valence orbitals, where the relativistic hopping and overlap integrals are calculated using the relativistic version of the Slater-Koster table. In the resulting energy-band structure, the appearance of the spin-orbit gap of 25 $\mu$eV at the $\emph{\textbf{k}}$ point due to the relativistic effect seen at low temperature, has also been shown by the theory.

\vspace{3mm}
\vspace{5mm}
\begin{multicols}{2}

\section*{INTRODUCTION}
\input{introduction}
\captionsetup[figure]{labelfont={bf},name={Fig.},labelsep=period}
\section*{THEORETICAL SECTION}
\input{Relativistic_TB}

\section*{METHOD}
\input{Method}

\section*{RESULT AND DISCUSSION}
\input{Result}
\section*{DATA AVAILABILITY}
The raw data required to reproduce these findings are not publicly available at this time but may be obtained from the author upon reasonable request.
\end{multicols}
\section*{APPENDIX}
\input{Appendix}
\begin{multicols}{2}
\bibliography{ref}
\bibliographystyle{ieeetr}
\end{multicols}

\end{document}

%% file: introduction.tex
 \paragraph{}
 The popularity in graphene research mainly stems from its unique and fascinating electronic \cite{novoselov2005two, bostwick2007quasiparticle, RevModPhys.81.109, chen2021electrically} and magnetic properties \cite{chen2021electrically,fan2019doping,PhysRevB.65.245420} and also due to the promise of potential applications and emergence of a new paradigm of relativistic-condensed matter physics \cite{geim2010rise}. Some of these physical properties are: it's minimum conductance never falls below the value corresponding to the smallest quantum unit of conductance even when the charge carrier concentrations tends to zero \cite{novoselov2005two}, tunable conductivity in twisted monolayer-bilayer graphene system \cite{chen2021electrically}, insulating behaviour at low temperatures \cite{chen2021electrically}, half-integer quantum hall effect \cite{PhysRevB.65.245420}, unconventional magnetic oscillations \cite{PhysRevB.69.075104} and strong orbital diamagnetism \cite{PhysRevB.91.214405}. The anomalous electronic properties of graphene arises from its characteristic band structure which is due to the effectively massless charge carriers with relativistic behaviour known as Dirac fermions at energies near Fermi level where the conical valance band and conduction band meet at a single point in momentum space called the Dirac points. 

Because of these physical properties of graphene it is desirable to investigate into a first-principle calculation method that takes into account both the relativistic effect and magnetic field. There are several studies where the electronic structure is calculated with non-relativistic tight binding model \cite{kundu2011tight,PhysRevB.66.035412} and a large number of these studies have been without considering the overlap integral \cite{PhysRevB.66.035412,PhysRevB.77.115410,RevModPhys.81.109}. There also exists a lot of credible relativistic calculation methods of electronic structures in absence of magnetic field \cite{Koelling_1975,takeda1978scalar,higuchi1995self} that has been successfully applied to f-electrons of rare earth compound $YbGa_2$ \cite{higuchi1995self} and compound $CeRu_2$ \cite{higuchi1996fermi}, but the electronic structure calculation method that can deal with magnetic field and relativistic effects simultaneously is desired for the study.

In 2015 a Non-perturbative Tight-Binding (TB) method called the magnetic-field-containing relativistic Tight-Binding approximation method (MFRTB method) was developed incorporating both the relativistic effects and magnetic field in the same footing and was applied to silicon crystal immersed in magnetic field \cite{PhysRevB.91.075122}. The MFRTB method was developed to alternatively describe the oscillatory behaviour of the magnetization as a function of magnetic field (the de Haas-van Alphen effect) from a view point of the first-principle calculation. The MFRTB method was successfully able to calculate the relativistic energy band structures of silicon in an external magnetic field \cite{PhysRevB.91.075122} that explained the suppression of softening in the elastic constant of boron-doped silicon in an external magnetic field, and also revisit the dHvA effect in a first-principle way \cite{PhysRevB.91.245101}. This method also presents the relativistic version of the Slater-Koster table in which the relativistic hopping integrals are explicitly expressed in terms of relativistic TB parameters for the interactions corresponding to $s$ and $p$ orbitals.

Our objective is to apply the MFRTB method to graphene. In order to apply the MFRTB method to materials in magnetic field, the magnetic hopping integrals must be expressed in terms of the relativistic hopping integrals for zero magnetic field and a magnetic field-dependent phase factor. So as an initial step, we need to formulate a relativistic TB model in zero magnetic field. The main objective of the study in this paper is to build a relativistic TB model for a regular hexagonal graphene lattice in absence of magnetic field whose results can be extended for calculations in presence of magnetic field. Starting from the Dirac equation for an electron in a periodic potential, the Dirac Hamiltonian matrix and overlap matrix of the system is constructed utilizing the relativistic Slater-Koster table considering the nearest neighbour interactions between $s$ and $p$ orbitals. Then by solving the general eigen value equation for different $\emph{\textbf{k}}$ points in the first Brillouin zone, the relativistic electronic structure of graphene is obtained. Our results have shown an appearance of a band gap of $25$ $\mu$eV for graphene at the Dirac points due to the spin-orbit coupling, which is in excellent agreement with the first principle calculation results of graphene using the linearized augmented plane-wave method incorporating the spin-orbit coupling \cite{PhysRevB.80.235431}. 

%% file: Relativistic_TB.tex
\begin{figure}[H]
    \centering
    \includegraphics[width = 7cm]{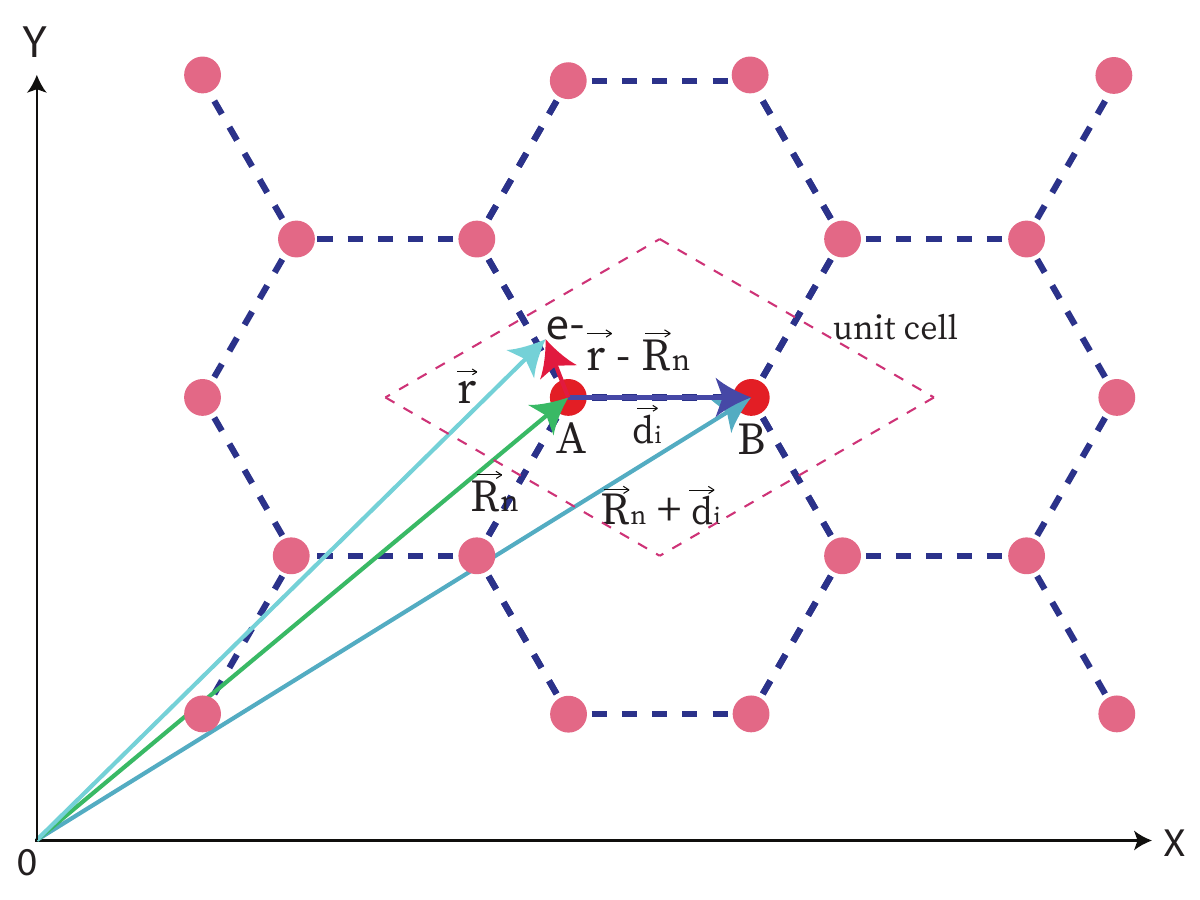}
    \caption{Illustration of honeycomb lattice and unit cell of graphene}
    \label{graphene lattice}
\end{figure}

\paragraph{}
Considering the two atoms of the unit cells of the honeycomb lattice, we proceed to establish the equations for the Hamiltonian matrix elements and the overlap matrix elements of the system. We represent the two atoms of the unit cell as `A' and `B'.
\subsection*{For lattice site `A'}
\paragraph{}
The Dirac equation for an electron at lattice site `A' is 
\begin{equation} \label{dirac}
    \hat{H_{A}}\Psi_{\alpha, \vec{k}}(\vec{r}) = E(\vec{k})\Psi_{\alpha, \vec{k}}(\vec{r}),
\end{equation}
with the atomic Hamiltonian at site `A' as
\begin{equation}
    \hat{H}_A = c\vec{\alpha} \cdot \vec{p} + \beta m c^2 + V_A(\vec{r} - \vec{R}_n).
\end{equation}
where $c$ denote the velocity of light, $m$ denote the electron rest mass, $\vec{\alpha} = (\alpha_{x},\alpha_{y},\alpha_{z})$, $\beta$ stans for the usual $4 \times 4$ matices \cite{leonard}, $V_{A}(\vec{r} - \vec{R_n})$ represents the Coulomb potential due to the nucleus of atom at site `A' located at $\vec{R}_n$, the subscripts $\alpha$ and $\vec{k}$ of the eigen function $\Psi_{\alpha, \vec{k}}(\vec{r})$ represents the band index and crystal momentum respectively. \\
Expanding $\Psi_{\alpha, \vec{k}}(\vec{r})$ by using the Bloch sum of relativistic atomic orbital $\phi_{nlJM}^{A}(\vec{r})$ of atom at `A' as the basis function, we have
\begin{equation}
    \Psi_{\alpha,\vec{k}}(\vec{r}) = \sum_{nlJM} C_{nlJM, \vec{K}}^A B_{nlJM}^A(\vec{r}),
\end{equation}
where $C_{nlJM, \vec{K}}^A$ is the expansion coefficient and $B_{nlJM}^{A,\vec{k}}(\vec{r})$ denote the Bloch sum which is written as

\begin{equation} \label{Bloch}
    B_{nlJM}^{A, \vec{k}}(\vec{r}) = \frac{1}{\sqrt{N}} \sum_{\vec{R_n}}e^{i\vec{k} \cdot \vec{R}_{n}} \phi_{nlJM}^A(\vec{r} - \vec{R_n}).
\end{equation}
The $n$, $l$, $J$ and $M$ in Eq. (\ref{Bloch}) are the principle, orbital angular momentum, total angular momentum and magnetic quantum numbers respectively. Here the relativistic atomic orbital obeys the following Dirac equation;
\begin{multline} 
    [c\vec{\alpha} \cdot \vec{p} + \beta m c^2 + V_A(\vec{r} - \vec{R}_n)]\phi_{nlJM}^A(\vec{r} - \vec{R_n}) \\= \varepsilon_{nlJ}^{A}\phi_{nlJM}^A(\vec{r} - \vec{R_n})
\end{multline}
where $\varepsilon_{nlJ}^{A}$ denotes the atomic spectrum for zero magnetic field at site `A', which is independent of $M$. Multiplying by $\phi_{n'l'J'M'}^{\dagger A'}(\vec{r} - \vec{R_{n'}})$ where $\vec{R_{n'}} \neq \vec{R_n}$, $A' \neq A$ and integrating and denoting the hopping and overlap integrals by $t_{n'l'J'M',nlJM}^{A'A}(\vec{R}_n - \vec{R}_{n'})$ and $s_{n'l'J'M',nlJM}^{A'A}(\vec{R}_n - \vec{R}_{n'})$ respectively, the hamiltonian and overlap matrix elements are defined as 
\begin{equation} \label{matrixH}
\begin{split}
    \textbf{H}_{n'l'J'M',nlJM}^{A'A}&(\vec{K}) \\
    &=  \sum_{\vec{R_n}}e^{i\vec{k} \cdot \vec{R_n}}t_{n'l'J'M',nlJM}^{A'A}(\vec{R_n}-\vec{R}_{n'}),
\end{split}
\end{equation}
and, 
\begin{equation} \label{matrixS}
\begin{split}
    \textbf{S}_{n'l'J'M',nlJM}^{A'A}&(\vec{K}) \\
    &= \sum_{\vec{R_n}}e^{i\vec{k} \cdot \vec{R_n}}s_{n'l'J'M',nlJM}^{A'A}(\vec{R}_n -\vec{R}_{n'}).
\end{split}
\end{equation}
So Eq. (\ref{dirac}) becomes 
\begin{equation} \label{sim}
\begin{split}
    \sum_{nlJM} &\textbf{H}_{n'l'J'M',nlJM}^{A'A}(\vec{K}) C_{nlJM,\vec{K}}^A = \\
    &E(\vec{K})\sum_{nlJM} \textbf{S}_{n'l'J'M',nlJM}^{A'A}(\vec{K})C_{nlJM,\vec{K}}^A.
\end{split}
\end{equation}

After evaluating the relativistic hopping and overlap integrals when the lattice site is considered to be at the origin ($\vec{R}_n = 0, d_i = d_j$) and when the lattice site is considered to be elsewhere ($\vec{R}_n \neq 0, d_i \neq d_j$) and by neglecting the three center integral, we get the matrix elements of the Hamiltonian and overlap Eq. (\ref{matrixH}) and Eq. (\ref{matrixS}) as:
\begin{equation} \label{HA}
    \begin{split}
        \textbf{H}&_{n'l'J'M',nlJM}^{A'A}(\vec{K}) = \\
        &(\varepsilon_{nlJ}^A+ \Delta\varepsilon_{nlJM}^A)\delta_{n'l'J'M', nlJM}\delta_{i,j}\\
        &+ \sum_{\vec{R}_{n}}e^{i\vec{k} \cdot \vec{R}_{n}}(1- \delta_{\vec{R}_{n},0}\delta_{i,j})t_{n'l'J'M',nlJM}^{new,A'A}(\vec{R}_{n}),
    \end{split}
\end{equation} 
with 
\begin{equation} \label{tnew}
\begin{split}
    t&_{n'l'J'M',nlJM}^{new,A'A}(\vec{R}_{n}) = \\
    &\frac{1}{2}(\varepsilon_{nlJ}^A+\varepsilon_{n'l'J'}^{A'})s_{n'l'J'M',nlJM}^{A'A}(\vec{r} - \vec{R}_{n}) \\
    &+ \int \phi_{n'l'J'M'}^{\dagger A'}(\vec{r})\left(\frac{V_{A'}(\vec{r}) + V_A(\vec{r}-\vec{{R}_{n}})}{2}\right)\\
    &\times \phi_{nlJM}^A(\vec{r} - \vec{{R}_{n}})d^3\vec{r},
\end{split}
\end{equation}
where $\Delta\varepsilon_{nlJM}^A$ is the crystal field energy due to the influence of other lattice sites on site `A' written as:

\begin{equation}
\begin{split}
    \Delta\varepsilon_{nlJM}^A =& \int \phi_{n'l'J'M'}^{\dagger A'}(\vec{r})\sum_{\vec{R}_{m} \neq\vec{R}_{n}}\sum_{A' \neq A}V_A(\vec{r}-\vec{R}_m)\\
    & \times \phi_{nlJM}^A(\vec{r}-\vec{R}_{m})d^3\vec{r}.
\end{split}
\end{equation}
And
\begin{multline} \label{SA}
    \textbf{S}_{n'l'J'M',nlJM}^{A'A}(\vec{K}) =  \delta_{n'l'J'M', nlJM}\delta_{i,j}\\
    + \sum_{\vec{R}_{n}}e^{i\vec{k} \cdot \vec{R}_{n}}(1-\delta_{\vec{R}_{n},0}\delta_{i,j})s_{n'l'J'M',nlJM}^{A'A}(\vec{R}_{n}),
\end{multline}
with 
\begin{multline}
    s_{n'l'J'M',nlJM}^{A'A}(\vec{R}_{n}) = \\
    \int \phi_{n'l'J'M'}^{\dagger A'}(\vec{r})\phi_{nlJM}^A(\vec{r} - \vec{{R}_{n}})d^3\vec{r}.
\end{multline}

\subsection*{For lattice site `B'}
\paragraph{}
Similarly for Lattice site `B' located at $\vec{R}_n + \vec{d}_i$ where $\vec{d}_i$ is the position vector from `A' to `B', we can formulate the Hamiltonian and overlap matrix elements as:
\begin{equation} \label{HB}
    \begin{split}
         \textbf{H}&_{n'l'J'M',nlJM}^{B'B}(\vec{K}) = \\
         &(\varepsilon_{nlJ}^B + \Delta\varepsilon_{nlJM}^B)\delta_{n'l'J'M', nlJM}\delta_{i,j} \\
         &+\sum_{\vec{R}_{n}} e^{i\vec{k} \cdot (\vec{R}_{n} + \vec{d}_i)}(1 - \delta_{\vec{R}_{n}, 0}\delta_{i,j})\\
         &\times t^{B'B, new}_{n'l'J'M',nlJM}(\vec{R}_{n} - \vec{d}_i + \vec{d}_j).
    \end{split}
\end{equation}
with
\begin{equation} 
    \begin{split}
        t&^{B'B, new}_{n'l'J'M',nlJM}(\vec{R}_{n} - \vec{d}_i + \vec{d}_j) = \\
        &\frac{1}{2} (\varepsilon_{n'l'J'}^{B'} + \varepsilon_{nlJ}^B) s_{n'l'J'M',nlJM}^{B'B}(\vec{r} - \vec{R}_{n} - \vec{d}_i + \vec{d}_j)\\
        &+\int \phi_{n'l'J'M'}^{\dagger B'}(\vec{r})
        \frac{V_{B'}(\vec{r}) + V_B(\vec{r} - \vec{R}_{n} - \vec{d}_i + \vec{d}_j)}{2}\\
        &\times \phi_{nlJM}^B(\vec{r} - \vec{R}_{n} - \vec{d}_i + \vec{d}_j)d^3\vec{r},
    \end{split}
\end{equation}
where $\Delta\varepsilon_{nlJM}^B$ is the crystal field energy  due to the influence of other lattice sites on site `B', written as:
\begin{equation}
    \begin{split}
        \Delta\varepsilon_{nlJM}^B = &\int \phi_{n'l'J'M'}^{\dagger B'}(\vec{r}) \sum_{\vec{R}_m \neq \vec{R}_{n}} \sum_{B' \neq B} V_{B'}(\vec{r} - \vec{R}_m - \vec{d}_j)\\
        & \times \phi_{nlJM}^B(\vec{r} - \vec{R}_m - \vec{d}_j)d^3\vec{r}.
    \end{split}
\end{equation}
And
\begin{equation} \label{SB}
\begin{split}
    \textbf{S}&_{n'l'J'M', nlJM}^{B'B}(\vec{K}) = \delta_{n'l'J'M', nlJM}\delta_{i,j} \\
    &+  \sum_{\vec{R}_{n}}e^{i\vec{k} \cdot (\vec{R}_{n} + \vec{d}_i)}(1-\delta_{\vec{R}_{n},0}\delta_{i,j})\\
    &\times s_{n'l'J'M',nlJM}(\vec{R}_{n} - \vec{d}_i + \vec{d}_j)
\end{split}
\end{equation}
with
\begin{multline}
    s_{n'l'J'M',nlJM}^{B'B}(\vec{R}_{n} - \vec{d}_i + \vec{d}_j) \\
    = \int \phi_{n'l'J'M'}^{\dagger B'}(\vec{r}) \phi_{nlJM}^{B}(\vec{r} - \vec{R}_{n} - \vec{d}_i + \vec{d}_j) d^3\vec{r}.
\end{multline}
\subsection*{Hamiltonian and overlap matrix elements of the system}
\paragraph{}
The equations of Hamiltonian matrix elements for site `A' and `B' differs only in the position of lattice sites, so formulating a general form of the Hamiltonian and overlap matrix elements considering the nearest neighbouring interactions in a hexagonal lattice, 
\begin{equation} \label{Hamiltonian}
    \begin{split}
       \textbf{H}&_{n'l'J'M',nlJM}(\vec{K}) = \\ &(\varepsilon_{nlJ}+ \Delta\varepsilon_{nlJM})\delta_{n'l'J'M', nlJM}\delta_{i,j}\\
       &+(1 - \delta_{\vec{R}_{n}, 0}\delta_{i,j})\sum_{\vec{R}_{n}} \sum\limits_{i = 1}^3  e^{i\vec{k} \cdot (\vec{R}_{n} + \vec{d}_i)}\\
       &\times t^{ new}_{n'l'J'M',nlJM}(\vec{R}_{n} - \vec{d}_i + \vec{d}_j).
    \end{split}
\end{equation}
and 
\begin{equation} \label{Overlap}
\begin{split}
     \textbf{S}_{n'l'J'M', nlJM}&(\vec{K}) =  \delta_{n'l'J'M', nlJM}\delta_{i,j}\\
     &+(1-\delta_{\vec{R}_{n},0}\delta_{i,j})\sum_{\vec{R}_{n''}}\sum\limits_{i = 1}^3 e^{i\vec{k} \cdot (\vec{R}_{n} + \vec{d}_i)}\\
     &\times s_{n'l'J'M',nlJM}(\vec{R}_{n} - \vec{d}_i + \vec{d}_j).
\end{split}
\end{equation}
Here the values of $i$ goes from 1 to 3 because there are 3 nearest neighbours for every atom in a honeycomb lattice of graphene with $\vec{d}_i$ as their position vectors. The superscripts ‘A’ and ‘B’ are eliminated because both are carbon atoms. Then Eq.(\ref{sim}) is rewritten as:
\begin{equation} \label{simultaneous4graphene}
    \begin{split}
        \sum_{nlJM}&[(\varepsilon_{nlJ} + \Delta \varepsilon_{nlJM} )\delta_{n'l'J'M',nlJM}\\
        &+\sum_{\vec{R}_{n''}} \sum\limits_{i = 1}^3e^{i \vec{k} \cdot (\vec{R}_{n} + \vec{d}_i)}(1 - \delta_{\vec{R}_{n}, 0}\delta_{i,j})\\
        &\times t_{n'l'J'M',nlJM}^{new}(\vec{R}_{n} - \vec{d}_i + \vec{d}_j)]C_{nlJM,\vec{k}}^i\\
        & = E(\vec{k}) [\delta_{n'l'J'M', nlJM}\delta_{i,j}\\
        &+\sum_{\vec{R}_{n}} \sum\limits_{i = 1}^3e^{i \vec{k} \cdot (\vec{R}_{n} + \vec{d}_i)}(1 - \delta_{\vec{R}_{n}, 0}\delta_{i,j})\\
        &\times s_{n'l'J'M',nlJM}^{ij}(\vec{R}_{n} - \vec{d}_i + \vec{d}_j)] C_{nlJM,\vec{k}}^i
    \end{split}
\end{equation}
Eq. (\ref{simultaneous4graphene}) is called the simultaneous equation which is to be solved to obtain the energy eigen values $E(\vec{k})$.

%% file: Method.tex
\subsection*{Construction of Hamiltonian and overlap matrix of the system}
\paragraph{}
First, Eq. (\ref{Hamiltonian}) and Eq. (\ref{Overlap}) are applied to carbon atoms arranged in hexagonal lattice with taking into account the hopping and overlap integrals between the outermost shells of the nearest-neighbouring atoms. Namely the following eight atomic states in the $2s$ and $2p$ orbitals of each carbon atoms are taken into consideration:
$(2, 0, \frac{1}{2}, \frac{1}{2})$, $(2, 0, \frac{1}{2}, -\frac{1}{2})$, $(2, 1, \frac{1}{2}, \frac{1}{2})$, $(2, 1, \frac{3}{2}, \frac{1}{2})$, $(2, 1, \frac{1}{2}, -\frac{1}{2})$, $(2, 1, \frac{3}{2}, -\frac{1}{2})$, $(2, 1, \frac{3}{2}, \frac{3}{2})$ and $(2, 1, \frac{3}{2}, -\frac{3}{2})$. To obtain the numerical values of $t_{n'l'J'M',nlJM}^{new}(\vec{R}_{n} - \vec{d}_i + \vec{d}_j)$ and $s_{n'l'J'M',nlJM}^{new}(\vec{R}_{n} - \vec{d}_i + \vec{d}_j)$ we used the relativistic version of the Slater-Koster table \cite{PhysRevB.91.075122} to express the hopping and overlap integrals in terms of the TB parameters $K_{d}^{i,j}(n'l'J',nlJ)_{|M|}$ and $S_{d}^{i,j}(n'l'J',nlJ)_{|M|}$ and then solved Eq. (\ref{tnew}) and Eq. (\ref{SA}). Since the crystal field energy is very small compared to the atomic spectrum, we have neglected $\Delta\varepsilon_{nlJM}$ and used the atomic spectrum $\varepsilon_{nlJ}$ that is calculated using density functional theory \cite{PhysRev.136.B864, PhysRev.140.A1133} based on local density approximation \cite{higuchi2019reduced}. The numerical values of the term $\varepsilon_{nlJ} + \Delta\varepsilon_{nlJM}$ are listed in Table (\ref{electronic spectra}). The numerical values of the TB parameters for the 8 atomic states of the nearest neighbouring atoms of graphene $K_{1}(n'l'J',nlJ)_{|M|}$ and $S_{1}(n'l'J',nlJ)_{|M|}$ has been calculated \cite{higuchi2019reduced} which are listed in Table (\ref{TB_parameters_values}).

\begin{table}[H]
    \centering
    \small
    \caption{\textbf{Numerical values of the sum of electronic spectra and crystal field energy\cite{higuchi2019reduced}.}}
    \begin{tabular}{|c|c|}
        \hline
        Energy States & Numerical values (eV)\\
        \hline
        $\varepsilon_{20\frac{1}{2}}+ \Delta\varepsilon_{20\frac{1}{2}M}$ & -8.37 \\
        
        $\varepsilon_{21\frac{1}{2}}+ \Delta\varepsilon_{21\frac{1}{2}M}$ & 0 \\
        
        $\varepsilon_{21\frac{3}{2}}+ \Delta\varepsilon_{21\frac{3}{2}M}$ & $8.305 \times 10^{-3}$\\
        \hline
    \end{tabular}
    \label{electronic spectra}
\end{table}

\begin{table} [H]
    \centering
    \caption{\textbf{Relativistic TB parameters for the hopping and overlap integrals for nearest neighbouring graphene atoms.}}
    \begin{tabular}{|c|c|}
    \hline
        $(nlJM)$ & Numerical values (eV) \\
        \hline
        $K_1(20\frac{1}{2},20\frac{1}{2})_{\frac{1}{2}}$ & -5.727 \\
        $K_1(20\frac{1}{2},21\frac{1}{2})_{\frac{1}{2}}$ & -3.226 \\
        $K_1(20\frac{1}{2},21\frac{3}{2})_{\frac{1}{2}}$ & 4.587\\
        $K_1(21\frac{1}{2},21\frac{1}{2})_{\frac{1}{2}}$ & -$1.81 \times 10^{-2}$\\
        $K_1(21\frac{1}{2},21\frac{3}{2})_{\frac{1}{2}}$ & -4.298\\
        $K_1(21\frac{3}{2},21\frac{3}{2})_{\frac{1}{2}}$ & 3.01\\
        $K_1(21\frac{3}{2},21\frac{3}{2})_{\frac{3}{2}}$ & -3.064\\
        $S_1(20\frac{1}{2},20\frac{1}{2})_{\frac{1}{2}}$ & $1.012 \times 10^{-1} $\\
        $S_1(20\frac{1}{2},21\frac{1}{2})_{\frac{1}{2}}$ & $9.739 \times 10^{-2}$\\
        $S_1(20\frac{1}{2},21\frac{3}{2})_{\frac{1}{2}}$ & $-1.392 \times 10^{-1}$\\
        $S_1(21\frac{1}{2},21\frac{1}{2})_{\frac{1}{2}}$ & $-7.904 \times 10^{-2}$\\
        $S_1(21\frac{1}{2},21\frac{3}{2})_{\frac{1}{2}}$ & $2.081 \times 10^{-1}$\\
        $S_1(21\frac{3}{2},21\frac{3}{2})_{\frac{1}{2}}$ & $-2.289 \times 10^{-1}$\\
        $S_1(21\frac{3}{2},21\frac{3}{2})_{\frac{3}{2}}$ & $6.802 \times 10^{-2} $\\
        \hline
    \end{tabular}
    
    \label{TB_parameters_values}
\end{table}
\paragraph{}
 To solve Eq. (\ref{simultaneous4graphene}) we first constructed a $16 \times 16$ Hamiltonian and overlap matrices using the values from Table (\ref{electronic spectra}) and Table (\ref{TB_parameters_values}) and using Eq. (\ref{Hamiltonian}) and Eq. (\ref{Overlap}) for the 8 atomic states at sites `$A$', `$B$' and `$A'$', `$B'$'. The first term vanishes for the non-diagonal diagonal elements (since $n'l'J'M' \neq nlJM$ and $i \neq j$) and the second term vanishes for the diagonal elements (since $i = j$ and $\vec{R}_n = 0$). The tabular representation of thus-constructed Hamiltonian and overlap matrix is shown in the Appendix section.
 \subsection*{Electronic structure calculation}
 \begin{figure}[H]
    \centering
    \begin{subfigure}[b]{0.45\textwidth}
        \centering
        \includegraphics[width = 6.5cm]{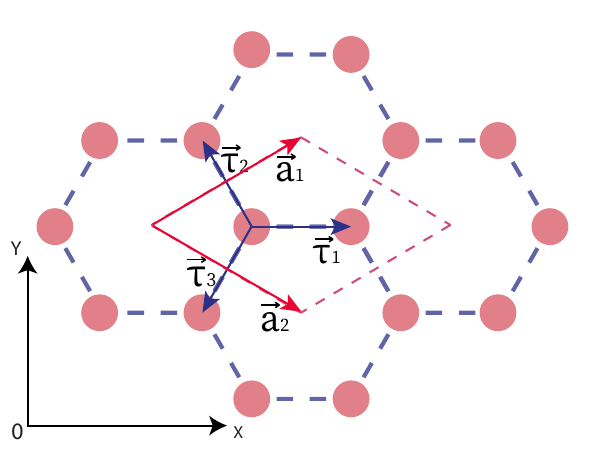}
        \caption{Illustration of real lattice and unit cell of graphene.}
        \label{real lattice}
    \end{subfigure}
    \hfill
    \begin{subfigure}[b]{0.45\textwidth}
        \centering
        \includegraphics[width = 6.5cm]{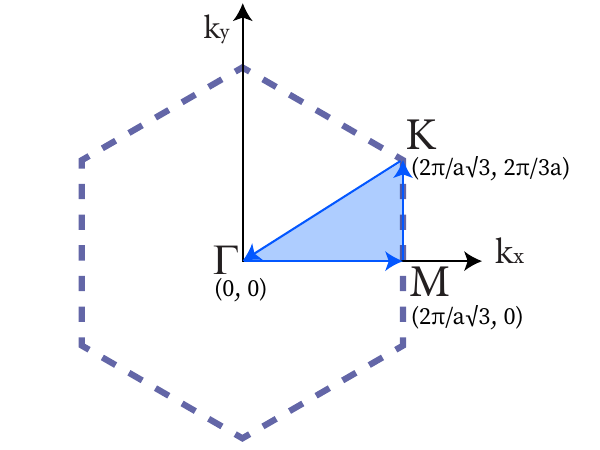}
        \caption{Illustration of reciprocal lattice and first Brillouin zone of graphene.}
        \label{reciprocal lattice}
    \end{subfigure}
\caption{Illustration of graphene lattice in real and reciprocal space.}
\label{real_reciprocal_lattice}
\end{figure}
\paragraph{}
Considering an atom to be at the origin, the coordinates of its three nearest neighbours are $(\frac{a}{\sqrt{3}}, 0)$, $(-\frac{a}{\sqrt{3}}, \frac{a}{2})$ and $(-\frac{a}{\sqrt{3}}, -\frac{a}{2})$ where $a = 2.4 \AA$ is the lattice constant. It can be shown that the position coordinates of the high symmetry points in the reciprocal lattice are as follows:
\begin{equation}
\begin{split}
    &\Gamma = (0,0)\\
    &K = \left(\frac{2\pi}{a\sqrt{3}}, \frac{2\pi}{3a}\right)\\
    &M = \left(\frac{2\pi}{a\sqrt{3}}, 0\right).
\end{split}    
\end{equation}

After constructing the Hamiltonian and overlap matrix, we solved the general eigen value Eq. (\ref{sim}) along a closed contour $K-\Gamma-M-K$ in the $\vec{k}$ space to get the energy eigen values $E(\Vec{k})$ and obtained the electronic band structure.

%% file: Result.tex
\begin{figure}[H]
        \centering
        \includegraphics[width = 8cm]{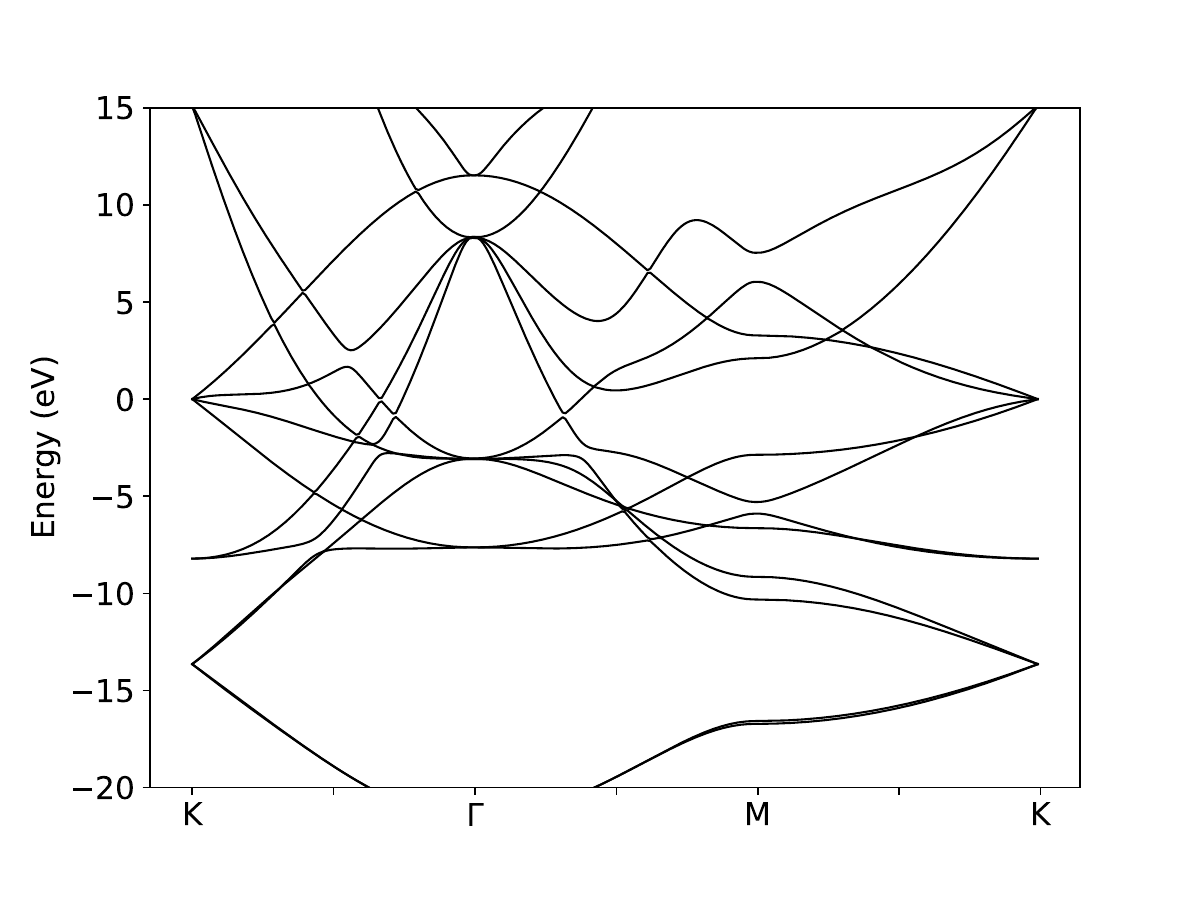}
        \caption{Enlarged view of the energy bands. The distinct $\pi\pi*$ and $\sigma \sigma*$ bonds are successfully replicated.}
        \label{croped_band}
\end{figure}

\paragraph{}
The energy diagram obtained from the calculation along the closed contour $K-\Gamma-M-K$ is shown in Fig.(\ref{croped_band}). The obtained band gap energy at the $K$ point was $2.58 \times 10^{-5}$ eV whose value is in close agreement with the value shown by the first-principle calculations considering spin-orbit interaction done by the FLAPW method \cite{PhysRevB.80.235431}. As we have considered the atomic states of the valance orbitals as the standard basis for constructing the Hamiltonian and overlap matrix, the energy bands corresponding to the $\sigma$ and $\pi$ bonds, namely the $\sigma \sigma^*$ band formed by the electrons in $2s$, $2p_x$ and $2p_y$ orbitals and $\pi \pi*$ band formed by electrons in the $2p_z$ orbital have been correctly reproduced. The $\pi$ band is very much similar to the original results of the non-relativistic electronic structure calculation of graphene produced by the energy dispersion relation (Eq. 3) in Ref. \cite{saito1992electronic}. An importance difference in our method is the appearance of a band gap of $25 \mu$eV due to the relativistic effect, particularly the spin-orbit coupling. Evidently, the relativistic effects is not so significant in graphene as the spin-orbit coupling is relatively week in carbon. However, for the studies regarding electronic and magnetic properties of heavier element materials, the need for a relativistic model becomes important as the spin-orbit coupling becomes larger and the relativistic effects becomes significant. 

In this paper we have only considered the nearest neighbouring interaction of the valence orbitals of carbon atoms. This model can very well be extended to include the second or third nearest neighbouring interactions which would give additional terms in the Hamiltonian matrix element (Eq. \ref{Hamiltonian}) and overlap matrix element Eq. (\ref{Overlap}). This model can easily be extended to other system in hexagonal lattice like silicene for example, by simply replacing the TB parameters (Table \ref{electronic spectra} and \ref{TB_parameters_values}) with the TB parameters of Si atoms in hexagonal lattice. This model can also be applied to other homogeneous or heterogeneous systems by calculating the relativistic TB parameters $K_{d}^{i,j}(n'l'J',nlJ)_{|M|}$ and $S_{d}^{i,j}(n'l'J',nlJ)_{|M|}$ for the respective atoms by solving the Spherical harmonics and using the relativistic Stater-Koster table \cite{PhysRevB.91.075122} to express the hopping and overlap integrals in terms of the TB parameters. 

Through this work, we have validated the success of novel, semi-empirical method called the MFRTB method to calculate the electronic band structure of graphene taking relativistic effects into consideration. The success of the method is demonstrated by correctly predicting the splitting of bands at the Dirac $K$ point. The magnetic field effects can also be incorporated into the present model by utilizing magnetic Bloch theorem and approximating the magnetic hopping integrals as the relativistic hopping integrals multiplied by the magnetic-field-dependent phase factor.
\section*{CONCLUSION}
In conclusion, we have generated a relativistic Tight-Binding model for carbon atoms in a regular hexagonal graphene lattice. Constructing the Dirac Hamiltonian of the system, we have calculated the relativistic electronic band structure of graphene which has accurately predicted the spin-orbit band gap. The results of this study can be extended for the calculations in presence of magnetic field to analyse the magnetic field effects. This is a coherent relativistic Tight-Binding model that can be applied to accurately predict the relativistic effects for various systems.
\section*{ACKNOWLEDGEMENT}
\input{Ack}

%% file: Ack.tex
\paragraph{}
This work was supported by the Department of Physics, Kathmandu University, Dhulikhel, Nepal. 

%% file: Appendix.tex
\begin{table} [H]
\begin{center}
\footnotesize
\caption{Hamiltonian matrix representation in tabular form.}
\scalebox{0.55}{
    \begin{tabular}{c c c c c c c c c | c c c c c c c c}
         & & & & $A$ & & & & & & & & & $B$ & & & \\
         $\times$ & $(2, 0, \frac{1}{2}, \frac{1}{2})$ &  $(2, 0, \frac{1}{2}, -\frac{1}{2})$ & $(2, 1, \frac{1}{2}, \frac{1}{2})$ & $(2, 1, \frac{1}{2}, -\frac{1}{2})$ & $(2, 1, \frac{3}{2}, \frac{3}{2})$ & $(2, 1, \frac{3}{2}, \frac{1}{2})$ & $(2, 1, \frac{3}{2}, -\frac{1}{2})$ & $(2, 1, \frac{3}{2}, -\frac{3}{2})$ & $(2, 0, \frac{1}{2}, \frac{1}{2})$ &  $(2, 0, \frac{1}{2}, -\frac{1}{2})$ & $(2, 1, \frac{1}{2}, \frac{1}{2})$ & $(2, 1, \frac{1}{2}, -\frac{1}{2})$ & $(2, 1, \frac{3}{2}, \frac{3}{2})$ & $(2, 1, \frac{3}{2}, \frac{1}{2})$ & $(2, 1, \frac{3}{2}, -\frac{1}{2})$ & $(2, 1, \frac{3}{2}, -\frac{3}{2})$ \\
        $(2, 0, \frac{1}{2}, \frac{1}{2})$ & $a_{1,1}$ & $0$ & $0$ & $0$ & $0$ & $0$ & $0$ & $0$ & $a_{1,9}$ & $a_{1,10}$ & $a_{1,11}$ & $a_{1,12}$ & $a_{1,13}$ & $a_{1,14}$ & $a_{1,15}$ & $a_{1,16}$ \\
        $(2, 0, \frac{1}{2}, -\frac{1}{2})$ & $0$ & $a_{2,2}$ & $0$ & $0$ & $0$ & $0$ & $0$ & $0$ & $a_{2,9}$ & $a_{2,10}$ & $a_{2,11}$ & $a_{2,12}$ & $a_{2,13}$ & $a_{2,14}$ & $a_{2,15}$ & $a_{2,16}$ \\
        $(2, 1, \frac{1}{2}, \frac{1}{2})$ & $0$ & $0$ & $a_{3,3}$ & $0$ & $0$ & $0$ & $0$ & $0$ & $a_{3,9}$ & $a_{3,10}$ & $a_{3,11}$ & $a_{3,12}$ & $a_{3,13}$ & $a_{3,14}$ & $a_{3,15}$ & $a_{3,16}$ \\
        $(2, 1, \frac{1}{2}, -\frac{1}{2})$ & $0$ & $0$ & $0$ & $a_{4,4}$ & $0$ & $0$ & $0$ & $0$ & $a_{4,9}$ & $a_{4,10}$ & $a_{4,11}$ & $a_{4,12}$ & $a_{4,13}$ & $a_{4,14}$ & $a_{4,15}$ & $a_{4,16}$ \\
        $(2, 1, \frac{3}{2}, \frac{3}{2})$ & $0$ & $0$ & $0$ & $0$ & $a_{5,5}$ & 0 & 0 & 0 & $a_{5,9}$ & $a_{5,10}$ & $a_{5,11}$ & $a_{5,12}$ & $a_{5,13}$ & $a_{5,14}$ & $a_{5,15}$ & $a_{5,16}$ \\
        $(2, 1, \frac{3}{2}, \frac{1}{2})$ & $0$ & $0$ & $0$ & $0$ & $0$ & $a_{6,6}$ & $0$ & $0$ & $a_{6,9}$ & $a_{6,10}$ & $a_{6,11}$ & $a_{6,12}$ & $a_{6,13}$ & $a_{6,14}$ & $a_{6,15}$ & $a_{6,16}$ \\
        $(2, 1, \frac{3}{2}, -\frac{1}{2})$ & $0$ & $0$ & $0$ & $0$ & $0$ & $0$ & $a77$ & $0$ & $a_{7,9}$ & $a_{7,10}$ & $a_{7,11}$ & $a_{7,12}$ & $a_{7,13}$ & $a_{7,14}$ & $a_{7,15}$ & $a_{7,16}$ \\
        $(2, 1, \frac{3}{2}, -\frac{3}{2})$ & $0$ & $0$ & $0$ & $0$ & $0$ & $0$ & $0$ & $a_{8,8}$ & $a_{8,9}$ & $a_{8,10}$ & $a_{8,11}$ & $a_{8,12}$ & $a_{8,13}$ & $a_{8,14}$ & $a_{8,15}$ & $a_{8,16}$ \\
         \hline
        $(2, 0, \frac{1}{2}, \frac{1}{2})$ & $a_{1,9}^*$ & $a_{2,9}^*$ & $a_{3,9}^*$ & $a_{4,9}^*$ & $a_{5,9}^*$ & $a_{6,9}^*$ & $a_{7,9}^*$ & $a_{8,9}^*$ & $a_{9,9}$ & $0$ & $0$ & $0$ & $0$ & $0$ & $0$ & $0$ \\
        $(2, 0, \frac{1}{2}, -\frac{1}{2})$ & $a_{1,10}^*$ & $a_{2,10}^*$ & $a_{3,10}^*$ & $a_{4,10}^*$ & $a_{5,10}^*$ & $a_{6,10}^*$ & $a_{7,10}^*$ & $a_{8,10}^*$ & $0$ & $a_{10,10}$ & $0$ & $0$ & $0$ & $0$ & $0$ & $0$ \\
        $(2, 1, \frac{1}{2}, \frac{1}{2})$ & $a_{1,11}^*$ & $a_{2,11}^*$ & $a_{3,11}^*$ & $a_{4,11}^*$ & $a_{5,11}^*$ & $a_{6,11}^*$ & $a_{7,11}^*$ & $a_{8,11}^*$ & $0$ & $0$ & $a_{11,11}$ & $0$ & $0$ & $0$ & $0$ & $0$ \\
        $(2, 1, \frac{1}{2}, -\frac{1}{2})$ & $a_{1,12}^*$ & $a_{2,12}^*$ & $a_{3,12}^*$ & $a_{4,12}^*$ & $a_{5,12}^*$ & $a_{6,12}^*$ & $a_{7,12}^*$ & $a_{8,12}^*$ & $0$ & $0$ & $0$ & $a_{12,12}$ & $0$ & $0$ & $0$ & $0$ \\
        $(2, 1, \frac{3}{2}, \frac{3}{2})$ & $a_{1,13}^*$ & $a_{2,13}^*$ & $a_{3,13}^*$ & $a_{4,13}^*$ & $a_{5,13}^*$ & $a_{6,13}^*$ & $a_{7,13}^*$ & $a_{8,13}^*$ & $0$ & $0$ & $0$ & $0$ & $a_{13,13}$ & $0$ & $0$ & $0$ \\
        $(2, 1, \frac{3}{2}, \frac{1}{2})$ & $a_{1,14}^*$ & $a_{2,14}^*$ & $a_{3,14}^*$ & $a_{4,14}^*$ & $a_{5,14}^*$ & $a_{6,14}^*$ & $a_{7,14}^*$ & $a_{8,14}^*$ & $0$ & $0$ & $0$ & $0$ & $0$ & $a_{14,14}$ & $0$ & $0$ \\
        $(2, 1, \frac{3}{2}, -\frac{1}{2})$ & $a_{1,15}^*$ & $a_{2,15}^*$ & $a_{3,15}^*$ & $a_{4,15}^*$ & $a_{5,15}^*$ & $a_{6,15}^*$ & $a_{7,15}^*$ & $a_{8,15}^*$ & $0$ & $0$ & $0$ & $0$ & $0$ & $0$ & $a_{15,15}$ & $0$ \\
        $(2, 1, \frac{3}{2}, -\frac{3}{2})$ & $a_{1,16}^*$ & $a_{2,16}^*$ & $a_{3,16}^*$ & $a_{4,16}^*$ & $a_{5,16}^*$ & $a_{6,16}^*$ & $a_{7,16}^*$ & $a_{8,16}^*$ & $0$ & $0$ & $0$ & $0$ & $0$ & $0$ & $0$ & $a_{16,16}$ \\
    \end{tabular}}
    \label{Htable}
\end{center}
\end{table}
where
\begin{equation*}
    a_{1,1} = \varepsilon_{20\frac{1}{2}}+ \Delta\varepsilon_{20\frac{1}{2}\frac{1}{2}}
\end{equation*}

\begin{equation*}
    \begin{split}
        a_{1,9} = &e^{i k_x \frac{a}{\sqrt{3}}}\left[\varepsilon_{20\frac{1}{2}}S\left(20\frac{1}{2}, 20\frac{1}{2}\right)_{\frac{1}{2}} + K\left(20\frac{1}{2}, 20\frac{1}{2}\right)_{ \frac{1}{2}}\right]\\
        &+e^{-i k_x \frac{a}{2\sqrt{3}} + i k_y \frac{a}{2}}\left[\varepsilon_{20\frac{1}{2}}S\left(20\frac{1}{2}, 20\frac{1}{2}\right)_{\frac{1}{2}} + K\left(20\frac{1}{2}, 20\frac{1}{2}\right)_{ \frac{1}{2}}\right]\\
        &+e^{-i k_x \frac{a}{2\sqrt{3}} - i k_y \frac{a}{2}}\left[\varepsilon_{20\frac{1}{2}}S\left(20\frac{1}{2}, 20\frac{1}{2}\right)_{\frac{1}{2}} + K\left(20\frac{1}{2}, 20\frac{1}{2}\right)_{ \frac{1}{2}}\right]
    \end{split}
\end{equation*}

\begin{equation*}
    a_{1,10} = 0
\end{equation*}
\begin{equation*}
    a_{1,11} = 0
\end{equation*}
and so on... 
\begin{table} [H]
\begin{center}
\footnotesize
\caption{Overlap matrix representation in tabular form.}
\scalebox{0.55}{
    \begin{tabular}{c c c c c c c c c | c c c c c c c c}
         & & & & $A$ & & & & & & & & & $B$ & & & \\
         $\times$ & $(2, 0, \frac{1}{2}, \frac{1}{2})$ &  $(2, 0, \frac{1}{2}, -\frac{1}{2})$ & $(2, 1, \frac{1}{2}, \frac{1}{2})$ & $(2, 1, \frac{1}{2}, -\frac{1}{2})$ & $(2, 1, \frac{3}{2}, \frac{3}{2})$ & $(2, 1, \frac{3}{2}, \frac{1}{2})$ & $(2, 1, \frac{3}{2}, -\frac{1}{2})$ & $(2, 1, \frac{3}{2}, -\frac{3}{2})$ & $(2, 0, \frac{1}{2}, \frac{1}{2})$ &  $(2, 0, \frac{1}{2}, -\frac{1}{2})$ & $(2, 1, \frac{1}{2}, \frac{1}{2})$ & $(2, 1, \frac{1}{2}, -\frac{1}{2})$ & $(2, 1, \frac{3}{2}, \frac{3}{2})$ & $(2, 1, \frac{3}{2}, \frac{1}{2})$ & $(2, 1, \frac{3}{2}, -\frac{1}{2})$ & $(2, 1, \frac{3}{2}, -\frac{3}{2})$ \\
        $(2, 0, \frac{1}{2}, \frac{1}{2})$ & 1 & 0 & 0 & 0 & 0 & 0 & 0 & 0 & $b_{1,9}$ & $b_{1,10}$ & $b_{1,11}$ & $b_{1,12}$ & $b_{1,13}$ & $b_{1,14}$ & $b_{1,15}$ & $b_{1,16}$ \\
        $(2, 0, \frac{1}{2}, -\frac{1}{2})$ & 0 & 1 & 0 & 0 & 0 & 0 & 0 & 0 & $b_{2,9}$ & $b_{2,10}$ & $b_{2,11}$ & $b_{2,12}$ & $b_{2,13}$ & $b_{2,14}$ & $b_{2,15}$ & $b_{2,16}$ \\
        $(2, 1, \frac{1}{2}, \frac{1}{2})$ & 0 & 0 & 1 & 0 & 0 & 0 & 0 & 0 & $b_{3,9}$ & $b_{3,10}$ & $b_{3,11}$ & $b_{3,12}$ & $b_{3,13}$ & $b_{3,14}$ & $b_{3,15}$ & $b_{3,16}$ \\
        $(2, 1, \frac{1}{2}, -\frac{1}{2})$ & 0 & 0 & 0 & 1 & 0 & 0 & 0 & 0 & $b_{4,9}$ & $b_{4,10}$ & $b_{4,11}$ & $b_{4,12}$ & $b_{4,13}$ & $b_{4,14}$ & $b_{4,15}$ & $b_{4,16}$\\
        $(2, 1, \frac{3}{2}, \frac{3}{2})$ & 0 & 0 & 0 & 0 & 1 & 0 & 0 & 0 & $b_{5,9}$ & $b_{5,10}$ & $b_{5,11}$ & $b_{5,12}$ & $b_{5,13}$ & $b_{5,14}$ & $b_{5,15}$ & $b_{5,16}$ \\
        $(2, 1, \frac{3}{2}, \frac{1}{2})$ & 0 & 0 & 0 & 0 & 0 & 1 & 0 & 0 & $b_{6,9}$ & $b_{6,10}$ & $b_{6,11}$ & $b_{6,12}$ & $b_{6,13}$ & $b_{6,14}$ & $b_{6,15}$ & $b_{6,16}$ \\
        $(2, 1, \frac{3}{2}, -\frac{1}{2})$ & 0 & 0 & 0 & 0 & 0 & 0 & 1 & 0 & $b_{7,9}$ & $b_{7,10}$ & $b_{7,11}$ & $b_{7,12}$ & $b_{7,13}$ & $b_{7,14}$ & $b_{7,15}$ & $b_{7,16}$ \\
        $(2, 1, \frac{3}{2}, -\frac{3}{2})$ & 0 & 0 & 0 & 0 & 0 & 0 & 0 & 1 & $b_{8,9}$ & $b_{8,10}$ & $b_{8,11}$ & $b_{8,12}$ & $b_{8,13}$ & $b_{8,14}$ & $b_{8,15}$ & $b_{8,16}$ \\
         \hline
        $(2, 0, \frac{1}{2}, \frac{1}{2})$ & $b_{1,9}^*$ & $b^*_{2,9}$ & $b_{3,9}^*$ & $b_{4,9}^*$ & $b_{5,9}^*$ & $b_{6,9}^*$ & $b_{7,9}^*$ & $b_{8,9}^*$ & 1 & 0 & 0 & 0 & 0 & 0 & 0 & 0 \\
        $(2, 0, \frac{1}{2}, -\frac{1}{2})$ & $b_{1,10}^*$ & $b_{2,10}^*$ & $b_{3,10}^*$ & $b_{4,10}^*$ & $b_{5,10}^*$ & $b_{6,10}^*$ & $b_{7,10}^*$ & $b_{8,10}^*$ & 0 & 1 & 0 & 0 & 0 & 0 & 0 & 0 \\
        $(2, 1, \frac{1}{2}, \frac{1}{2})$ & $b_{1,11}^*$ & $b_{2,11}^*$ & $b_{3,11}^*$ & $b_{4,11}^*$ & $b_{5,11}^*$ & $b_{6,11}^*$ & $b_{7,11}^*$ & $b_{8,11}^*$ & 0 & 0 & 1 & 0 & 0 & 0 & 0 & 0 \\
        $(2, 1, \frac{3}{2}, \frac{1}{2})$ & $b_{1,12}^*$ & $b_{2,12}^*$ & $b_{3,12}^*$ & $b_{4,12}^*$ & $b_{5,12}^*$ & $b_{6,12}^*$ & $b_{7,12}^*$ & $b_{8,12}^*$ & 0 & 0 & 0 & 1 & 0 & 0 & 0 & 0 \\
        $(2, 1, \frac{1}{2}, -\frac{1}{2})$ & $b_{1,13}^*$ & $b_{2,13}^*$ & $b_{3,13}^*$ & $b_{4,13}^*$ & $b_{5,13}^*$ & $b_{6,13}^*$ & $b_{7,13}^*$ & $b_{8,13}^*$ & 0 & 0 & 0 & 0 & 1 & 0 & 0 & 0 \\
        $(2, 1, \frac{3}{2}, -\frac{1}{2})$ & $b_{1,14}^*$ & $b_{2,14}^*$ & $b_{3,14}^*$ & $b_{4,14}^*$ & $b_{5,14}^*$ & $b_{6,14}^*$ & $b_{7,14}^*$ & $b_{8,14}^*$ & 0 & 0 & 0 & 0 & 0 & 1 & 0 & 0 \\
        $(2, 1, \frac{3}{2}, \frac{3}{2})$ & $b_{1,15}^*$ & $b_{2,15}^*$ & $b_{3,15}^*$ & $b_{4,15}^*$ & $b_{5,15}^*$ & $b_{6,15}^*$ & $b_{7,15}^*$ & $b_{8,15}^*$ & 0 & 0 & 0 & 0 & 0 & 0 & 1 & 0 \\
        $(2, 1, \frac{3}{2}, -\frac{3}{2})$ & $b_{1,16}^*$ & $b_{2,16}^*$ & $b_{3,16}^*$ & $b_{4,16}^*$ & $b_{5,16}^*$ & $b_{6,16}^*$ & $b_{7,16}^*$ & $b_{8,16}^*$ & 0 & 0 & 0 & 0 & 0 & 0 & 0 & 1 \\
    \end{tabular}}
    \label{Stable}
\end{center}
\end{table}
where
\begin{equation*}
    b_{1,9} = e^{ik_x\frac{a}{\sqrt{3}}}S\left(20\frac{1}{2},20\frac{1}{2}\right)_{\frac{1}{2}} + e^{-ik_x\frac{a}{2\sqrt{3}} + ik_y\frac{a}{2}}S\left(20\frac{1}{2},20\frac{1}{2}\right)_{\frac{1}{2}} + e^{-ik_x\frac{a}{2\sqrt{3}} - ik_y\frac{a}{2}}S\left(20\frac{1}{2},20\frac{1}{2}\right)_{\frac{1}{2}}
\end{equation*}

\begin{equation*}
    b_{1,10} = 0
\end{equation*}

\begin{equation*}
    b_{1,11} = 0
\end{equation*}

\begin{equation*}
\begin{split}
    b_{1,12} =& e^{ik_x\frac{a}{\sqrt{3}}}S\left(20\frac{1}{2},21\frac{1}{2}\right)_{\frac{1}{2}}\\
    &+ e^{-ik_x\frac{a}{2\sqrt{3}} + ik_y\frac{a}{2}}\left[\left(-\frac{1}{2}-i\frac{\sqrt{3}}{2}\right)S\left(20\frac{1}{2},21\frac{1}{2}\right)_{\frac{1}{2}}\right]\\
    &+ e^{-ik_x\frac{a}{2\sqrt{3}} - ik_y\frac{a}{2}}\left[\left(-\frac{1}{2}+i\frac{\sqrt{3}}{2}\right)S\left(20\frac{1}{2},21\frac{1}{2}\right)_{\frac{1}{2}}\right]
\end{split}
\end{equation*}
and so on...